\documentclass[conference]{IEEEtran}
\ifCLASSINFOpdf
\else
\fi
\usepackage[hyphens]{url}
\usepackage{multirow}
\usepackage[normalem]{ulem}
\usepackage{xspace}
\usepackage{graphicx}
\usepackage{tikz, pgfplots, pgfplotstable}
\usepackage{cryptocode}
\usepackage{subfigure}
\usepackage[htt]{hyphenat}

\newcommand{\acroc}{{\textit{CREAM}}\xspace} 

\newcommand\adv{\ensuremath{\sf{\mathcal Adv}}\xspace}

\usepackage{enumitem}
\newlist{myitemize}{itemize}{1}
\setlist[myitemize]{
    label=$\bullet$,
    align=left,
    leftmargin=*,
    nosep,
}

\begin{document}
%

\title{Towards Browser Controls to Protect Cookies from Malicious Extensions}


	

%
\author{\IEEEauthorblockN{Liam Tyler\IEEEauthorrefmark{1},
Ivan De Oliveira Nunes\IEEEauthorrefmark{2}}
\IEEEauthorblockA{\IEEEauthorrefmark{1}University of Zurich\\
Email: ltyler@ifi.uzh.ch}
\IEEEauthorblockA{\IEEEauthorrefmark{2}University of Zurich\\
Email: ivan.deoliveiranunes@uzh.ch}}



\maketitle

\begin{abstract}
Cookies are vital to the web's current operation, taking on several important and security-critical tasks such as user authentication. Due to their importance, browsers implement several cookie controls, such as the \texttt{Secure} and \texttt{HttpOnly} attributes to help protect cookies from malicious networks and websites. While effective, these controls overlook browser extensions: third-party HTML/JavaScript add-ons with access to privileged browser APIs and the ability to operate across multiple websites. As a result, malicious/compromised extensions can gain unrestricted access to a user's cookies.

In this work, we first analyze the prevalence of extensions that can access and modify cookies. Our analysis shows that these extensions reach hundreds of millions of users. We then propose a mechanism to protect cookies from malicious (or compromised) extensions, introducing two complementary cookie attributes: \textbf{BrowserOnly} and \textbf{Monitored}. The \textbf{BrowserOnly} attribute prevents extensions from accessing a cookie outright. However, not all cookies can be made inaccessible while preserving intended functionality. As such, the \textbf{Monitored} attribute serves as an alternative, maintaining accessibility while tying cookies to a single browser instance and logging any modifications made to the cookie by extensions. As a result, stolen \textbf{Monitored} cookies are unusable outside their original browser and servers can verify the logged modifications. To demonstrate the proposed concepts, we design and implement \acroc (\underline{C}ookie \underline{R}estrictions for \underline{E}xtension \underline{A}buse \underline{M}itigation) as a modified version of Chromium realizing these controls. Our evaluation shows effective cookie protection against malicious extensions and low run-time overheads.
\end{abstract}


%
\IEEEpeerreviewmaketitle

\section{Introduction}
\label{sec:intro}

Cookies remain a universal state management mechanism on the web. They allow servers to store data on the user's browser and for that data to be automatically returned in all subsequent requests \cite{rfc6265}. This enables the linking of related traffic into a session and the maintenance of data across it.

A common and security-critical use of cookies is user authentication, where cookies store an authenticated user session ID and replace credentials in future traffic~\cite{dacosta2012one, calzavara2017surviving}. 
This makes cookies a valuable target for attacks such as session hijacking, session fixation, and Cross-Site Request Forgery (CSRF). Session hijacking \cite{bugliesi2015cookiext, squarcina2023cookie, dacosta2012one} attempts to gain access to the user's account by stealing the value of their ``session cookie''. These attacks are not limited to session cookies alone as non-session cookies can still leak users' private information \cite{sivakorn2016cracked}. Similarly, session fixation \cite{calzavara2017surviving, squarcina2023cookie, johns2011reliable} injects a malicious session cookie with a known value into the user's browser. Once the user authenticates with the website, an adversary can use the known value to access the user's account. CSRF tricks the browser into issuing a cross-site request to perform some action as the user without their knowledge \cite{calzavara2017surviving, khodayari2022state, squarcina2023cookie}. 

Several cookie features have been introduced to combat these attacks \cite{mdn_use_cookie, rfc6265}. The \texttt{Secure} attribute restricts cookies to HTTPS traffic, ensuring they are always encrypted on the network. Similarly, the \texttt{HttpOnly} attribute prevents JavaScript (JS) from accessing cookies. Together, these attributes help mitigate session hijacking by preventing cookie theft from malicious JS and networks. The \texttt{SameSite} attribute restricts how cookies are attached to cross-site requests, reducing the threat of CSRF. Cookies also define two cookie prefixes: \texttt{\_\_Secure-} and \texttt{\_\_Host-}. These prefixes enforce that certain criteria are true when creating the cookie. As such, cookie prefixes help reduce the threat of session fixation by asserting how a cookie was created.

While effective against malicious websites and networks, the aforementioned controls do not protect against browser extensions. Extensions are third-party HTML/JS applications that add functionality to browsers and execute across multiple websites \cite{mdn_ext_2023, Chrome_ext}. As such, extensions can bypass several browser controls by default, including cookie controls \cite{chrome_cor_2012, agarwal2022helping}. For example, extensions can access \texttt{HTTPOnly} cookies despite being JS themselves. Similarly, extensions can access network traffic as it is processed in the browser. Thus, they can view and modify cookies within network headers despite TLS encryption, bypassing both the \texttt{Secure} and \texttt{SameSite} attributes. Further, in the current model, extensions can simply modify cookies to remove their security attributes altogether. Unsurprisingly, malicious/compromised extensions have been widely used in recent attacks~\cite{Lakshmanan_2025,Tal_2023,Toulas_2024} to steal cookies and gain access to user-authenticated sessions.
 
These permissive extension controls stem from the current browser trust model. Most existing browser controls assume two possible attackers, a web attacker \cite{barth2008security, chen2018we, roth2022security, reis2019site, carlini2012evaluation, barth2010protecting, squarcina2021can, agarwal2022spook} and a network attacker~\cite{calzavara2017surviving, chen2018we, roth2022security, carlini2012evaluation, barth2010protecting}. In both models, the attacker controls a malicious website, whereas a network attacker can also inspect and tamper with HTTP traffic. In both models, the attacker is an external party. Thus, everything inside the browser is considered trusted, including extensions. This is also reflected by the extension architecture itself assuming ``benign but buggy" extensions \cite{barth2010protecting}.

Instead of architectural controls, browser vendors try to prevent malicious extensions from becoming available in the first place. Before being publicly listed, extensions are reviewed to adhere to the browser's extension policies \cite{Kewisch_2024, chrome_ext_review_2021}. However, malicious extensions can still pass this review \cite{jagpal2015trends,kapravelos2014hulk,pantelaios2020you,Davis_2024}, requiring them to be pulled from users' browsers after the fact \cite{Toulas_2023, Kaya_Rickerd_2020}. Along with this, most extension permissions must be accepted at installation \cite{chrome_permsissions_2024}. Thus users can prevent malicious extension behavior by denying them permissions. However, users are often unaware of the risks associated with these permissions \cite{kariryaa2021understanding}. Regardless of either check, extensions can also ``become'' malicious. Browsers automatically update extensions in the background \cite{Wulf_2024}. As such, extensions can silently receive a malicious update while maintaining their current permissions \cite{pantelaios2020you}. Further, browsers allow installing extensions manually (typically for extension development) \cite{chrome_manual_install_2022}. This can be leveraged to bypass all checks by tricking the user into manually installing a malicious extension.

Prior work also demonstrates that vulnerable extensions can leak their permissions to a malicious website \cite{kim2023extending,carlini2012evaluation,yu2023coco}. Thus, compromised extensions can also be used to bypass existing cookie controls. To demonstrate this risk, we use the extension APIs to hijack test accounts on several websites (see Section \ref{sec:malicousness}). We also scrape 156,783 extensions from both Chrome's and Firefox's extension repositories to assess the prevalence of these APIs in current extensions. 

Motivated by these issues, this work aims to reevaluate the trust model employed by current browsers to mitigate risks posed by extensions. Prior work has sought to protect sessions from extensions by preventing their access to sensitive information on web pages~\cite{wang2021webenclave, vasiliadis2023writ, picazo2020after}. While these controls protect sessions from direct access by extensions, malicious extensions maintain access to session cookies. Thus adversaries can still gain unauthorized access to the session. 

\subsection*{\bf Contributions.}

We propose two new cookie attributes to protect cookies from malicious extensions: \texttt{BrowserOnly} and \texttt{Monitored}. Unlike previous controls, we consider untrusted and potentially malicious extensions (e.g., as in recent attacks of~\cite{Lakshmanan_2025,Tal_2023,Toulas_2024}). The \texttt{BrowserOnly} attribute prevents extensions from accessing cookies. While effective, blocking access to certain cookies would break some extension and website functionality. For these cases, the \texttt{Monitored} attribute allows extensions access to cookies but ties the cookies to a single browser instance and maintains a \texttt{changelog} for each cookie. When sending a \texttt{Monitored} cookie to a server, the \texttt{changelog} and other server-defined data is authenticated and sent alongside the cookie. As such, \texttt{Monitored} does not prevent direct cookie theft. Instead, it enables server-side detection of stolen and illegally modified cookies, preventing their subsequent use. Importantly, as discussed in Section~\ref{sec:privacy}, the proposed features \textit{do not imply privacy loss}. 


The \texttt{Monitored} attribute uses symmetric keys to link cookies to a single browser and authenticate their \texttt{chagelog}-s. Specifically, the browser generates a unique key for each site it visits and attaches the key to all outbound HTTPS requests to its respective site. The server then embeds this key as part of a cookie's \texttt{Monitored} attribute to identify the browser and verify the authenticated attribute in future requests. This verification ensures the cookie came from the correct browser and its \texttt{changelog} has not been tampered with. While these keys allow individual sites to identify the user's browser, they cannot be used for cross-site tracking as the browser uses a unique key for each site visited.

Enforcement of the \texttt{BrowserOnly} and \texttt{Monitored} attributes is obtained by modifying the browser architecture to revoke extensions' ability to access/modify these attributes before the underlying browser core. As a proof of concept, we implement these features as an open-source modified version of the Chromium browser~\cite{chromium} and evaluate our prototype concerning storage and run-time impacts.
In summary, our intended contributions are:

\begin{myitemize}
    \item We study Chrome's and Firefox's extension repositories to assess the prevalence of extensions that, if malicious or compromised, could threaten cookies.
    \item We propose a new browser trust model regarding extensions, in which we introduce two new cookie attributes to protect cookies from potentially malicious browser extensions. The \texttt{BrowserOnly} attribute prevents cookies from being accessed by extensions outright. The \texttt{Monitored} attribute leaves cookies accessible while allowing servers to detect cookie theft and tampering (allowing servers to refuse service to stolen cookies). To support this functionality, we design a mechanism to tie browser instances to cookies without linking users across domains.
    \item To realize this concept we implement and open-source \acroc \cite{anonymous642_cream} (\underline{C}ookie \underline{R}estrictions for \underline{E}xtension \underline{A}buse \underline{M}itigation) as a modified version of Chromium and evaluate its functionality and (modest) overheads.
\end{myitemize}

\section{Background}
\label{sec:background}

\subsection{Browser Architecture}
\label{sec:browser_arch}

\begin{figure}
    \centering
    \scalebox{.9}{
        
        \includegraphics[width=\columnwidth]{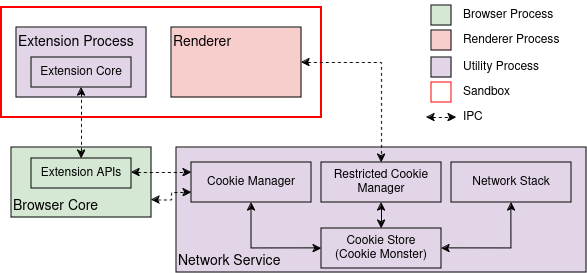}
    }
    \vspace{-.5em}
    \caption{The Browser Architecture and Cookie Interfaces}
    \vspace{-1em}
    \label{fig:cookie_man}
\end{figure}

Modern browsers implement a multi-process design \cite{chrome_proc_model,firefox_proc_model,safari_proc_model} (depicted in Figure \ref{fig:cookie_man}), leveraging OS-based inter-process isolation to provide strong separation of privilege within the browser. While different browsers have different implementations and naming conventions, all mainstream browsers follow a general model. Namely, they implement two main process types: browser and renderer processes \cite{chromium_multi, lim2021sok, barth2008security}. 

Renderer processes parse untrusted, potentially malicious, web content (i.e., HTML, CSS, JS) and build the final page displayed to the user \cite{chromium_multi, reis2019site, chromium_sandbox}. As such, they are unprivileged. Renderers also execute in a sandbox to further restrict their access to the underlying host. This forces all host-related requests to go through the browser process. 

The browser process is the main privileged process in the browser \cite{chromium_multi, barth2008security}. While there can be arbitrarily many renderer processes, each browser instance only has one browser process. The browser process interacts directly with the host, manages all other processes in the browser, and facilitates inter-process communication (IPC) between them~\cite{lim2021sok,kim2023extending}.

Browsers are also service-oriented \cite{Jesup_2017, chrome_servicification}. Many renderer and browser features have been moved to independent services that often execute within separate processes. For example, the browser's Network Service implements its network stack (all low-level networking code) and handles all network transactions \cite{chromium_net_serv, safari_proc_model, firefox_proc_types}. As these feature-specific processes have no standard name, we use Chrome's nomenclature of ``utility processes'' \cite{chromium_multi, lim2021sok} to refer to them. Utility processes can be privileged, unprivileged, sandboxed, or unsandboxed depending on their functionality.

In the remainder of this paper, when referencing ``the browser'' we are discussing the collection of features commonly referred to as the browser. This includes both the privileged browser process and utility processes, such as the Network Service. We refer to the browser process alone as the ``browser core'' instead. 

\subsection{Extension Architecture}
\label{sec:ext_arch}

All extensions have a \texttt{manifest.json} file \cite{Chrome_ext, barth2010protecting}. Currently, there are two manifest versions (V2 and V3), with slight differences in how attributes are defined, which extension APIs are available, and which browsers support them. This file lists the APIs and scripts an extension uses, websites it runs on, how it is loaded, resources it can access, and more. The browser requires the manifest to be present when an extension is installed and restricts the extension's behavior to what is stated in its manifest. This prevents extensions from gaining more privileges after installation, except for optional permissions which can be requested at any time, but must still be declared in the manifest file at installation.

An extension's functionality is divided into two components: the extension core and content scripts \cite{barth2010protecting, carlini2012evaluation}. The extension core is highly privileged with access to the extension APIs and the ability to connect to remote servers. To protect these privileges from untrusted websites, the extension core executes in its own process. Further, as different extensions are effectively independent programs, different extensions' cores cannot share the same process \cite{reis_moshchuk_2021}. However, despite being highly privileged, these processes are also run within a sandbox; again forcing all interactions with the underlying system through the browser \cite{barth2010protecting}.

This isolation prevents the extension core from interacting directly with any website. Instead, content scripts are injected into the page and have direct access to its content \cite{barth2010protecting, carlini2012evaluation}. These scripts execute within the renderer process of the website and are thus at an increased risk of attack. Due to this, content scripts are isolated from other scripts on the page \cite{barth2010protecting, kim2023extending}. Similarly, content scripts are unprivileged and cannot access most extension APIs. Instead, content scripts can only send messages to the extension core using the message-passing API. This aims to prevent a compromised renderer from gaining the powers of any extensions embedded within it.

\subsection {Cookies \& Cookie Management}
\label{sec:cookie_background}

At their simplest, cookies are key-value pairs associated with the server that creates them. Cookies can also set multiple optional attributes that further define how they are handled. Which server a cookie belongs to and what requests it should be attached to are determined by the \texttt{Domain} and \texttt{Path} attributes \cite{rfc6265}. These indicate which websites and which portion of those sites the cookie can be sent to respectively. If not specified, the browser uses the hostname and path of the website that created the cookie by default. Cookies can also be partitioned. Partitioning limits a cookie's accessibility to its embedding context such that cookies set by the same site but embedded on different pages are not accessible to each other \cite{chrome_chips_2023}. This is often used to prevent third-party cookies from tracking users across websites. As discussed in Section \ref{sec:intro}, cookies also have three security attributes: \texttt{HttpOnly}, \texttt{Secure}, and \texttt{SameSite}. They also support two prefixes that enforce certain cookie creation criteria \cite{mdn_use_cookie}. 
 
\subsubsection{Cookie Manager}
\label{sec:cookie_manager}

Cookies can be set in many ways. Most commonly, they are defined in the \texttt{Set-Cookie} HTTP header \cite{mdn_set-cookie} and created by the Network Service \cite{chromium_cookie_endpoints_2020}. As cookies provide state for network requests, they are stored within and are directly interactable from the Network Service. Cookies can also be created from a webpage through `document.cookie' \cite{mdn_document_cookie} or the Cookie Store API \cite{mdn_cookie_store_api}, by the browser core, and by extensions through the \texttt{Cookies} extension API \cite{Chrome_cookie_api}. However, none of these can interact with cookies directly. Instead, the Network Service exposes APIs for accessing its internal cookie store \cite{chromium_cookie_endpoints_2020}. The arrows in Figure \ref{fig:cookie_man} depict how different browser components access cookies. 

These interface APIs are implemented by the Cookie Manager and the Restricted Cookie Manager \cite{chromium_cookie_endpoints_2020}. The Cookie Manager allows unrestricted querying of the cookie store and is used by the browser core. Since the \texttt{Cookies} API is a browser-level API, extensions also access cookies through the Cookie Manager. The Restricted Cookie Manager, as the name implies, is more limited. The restricted manager can only access the cookies of a single specified origin. For this reason, the restricted manager is used by renderer processes to enforce stricter limitations on websites.

\subsubsection{Cookie Monster}
\label{sec:cookie_monster}

All cookies are maintained by the browser's cookie store. Chromium's concrete implementation of the cookie store is called the Cookie Monster \cite{chromium_cookie_monster}.
There are two ways to fetch cookies from the Cookie Monster \cite{chromium_cookie_code}. The Cookie Monster can either return all unexpired cookies or a subset of cookies that match some provided criteria. We refer to these methods as \texttt{GetAll} and \texttt{GetList} respectively. When using \texttt{GetAll}, the Cookie Monster returns all cookies within its store, regardless of their attributes. However, most fetches use \texttt{GetList} which takes a domain and a set of options, fetches all the provided domain's cookies, and filters them based on the provided options \cite{chromium_cookie_endpoints_2020}. These options include checks such as whether to allow \texttt{HttpOnly} or \texttt{Secure} cookies in the results.

To set a cookie (\texttt{Set} operation), the caller (browser core, renderer, extension, etc.) creates a new cookie and sends it to the Cookie Monster along with its domain and other cookie options \cite{chromium_cookie_code}. Upon receipt of the cookie, the Cookie Monster first checks that the cookie is valid. Similar to \texttt{GetList}, this validation uses the supplied options to ensure the cookie is properly configured (e.g., verifying JS is not setting an \texttt{HTTPOnly} cookie). If valid, the Cookie Monster then checks for any existing matching cookie. If a match is found, it is replaced with the new cookie. Otherwise, the cookie is added to the internal map of cookies as a new entry. 

\section{On the Extent of Extension Threats}
\label{sec:malicousness}

\subsection{Manipulating Cookies via Extension APIs}
\label{sec:case_study}

To illustrate the practical threats extensions pose to cookies, we implemented three test extensions targeting different extension APIs. Each extension was tested in Chrome against test accounts on several websites. Each test extension was built with both V2 and V3 manifest versions, when applicable. V2 extensions were recently deprecated in Chrome \cite{Li_2023}, however, at the time of testing they were still supported and V2 maintains support in other browsers \cite{Sullivan_2024}. We specifically target the \texttt{Cookies}, \texttt{WebRequest}, and \texttt{DeclarativeNetRequest} APIs, as they allow interaction with cookies. The \texttt{Cookies} API allows extensions to get, set, remove, and be notified of changes to cookies within the browser \cite{Chrome_cookie_api}. The cookies accessible to an extension are limited to those of websites it has permission to run on. Nonetheless, an extension can have permission to run on all websites (which is not uncommon). The \texttt{WebRequest} API allows extensions to view network traffic mid-processing \cite{chrome_web_req_2024}. As such, it allows extensions to access HTTP headers, including cookie headers. The \texttt{WebRequest} API can also be extended with the \texttt{WebRequestBlocking} permission to allow extensions to pause the execution of network events and edit request headers as well. Finally, the \texttt{DeclarativeNetRequest} API also allows extensions to modify network requests \cite{chrome_dec_2024, mdn_dec_net_2024}. However, unlike \texttt{WebRequest}, \texttt{DeclarativeNetRequest} cannot view the contents of the request. Each test extension targeted a single API as follows:

\begin{myitemize}

\item {\bf Case 1 - \texttt{Cookies} API:} This extension implements a cookie management interface allowing the user to view, edit, and delete browser cookies. When fetching cookies, the extension also forwards them to our test server. Using this, we captured the cookies for many websites including Google, Amazon, and Facebook. Adding the stolen cookies to a new Firefox instance allowed us to access the corresponding test accounts without their credentials. We note that these accesses require a combination of cookies rather than a single cookie, but as extensions can easily access all of a website's cookies this is not an obstacle.

\item {\bf Case 2 - \texttt{WebRequest} API:} This extension allows the user to intercept, view, and modify traffic for specified websites. When viewing the request, the request headers are also forwarded to our test server. From the headers, we can retrieve all the cookies sent to and set by the website. This extension can also modify request headers and thus edit the cookies sent to/received from the server. It is important to reiterate that modifying headers requires the \texttt{WebRequestBlocking} permission \cite{chrome_web_req_2024} which is only available for V2 extensions in Chrome. While V3 extensions cannot edit requests they can still steal cookies.

\item {\bf Case 3 - \texttt{DeclarativeNetRequest} API:} This extension allows the user to create rules that modify network traffic without viewing its content. As such, it allows us to edit the cookies in traffic, however, as the extension cannot inspect the request headers, it cannot steal cookies. The Declarative API  only exists for V3 extensions.

\end{myitemize}

\begin{table*}   
\centering
    \caption{Breakdown of extensions versioning, host permissions, and access to cookies}
    \vspace{-.5em}
    \scalebox{.75}{  
            \begin{tabular}{|c||c|c||c|c|c||c|c|c|c|}
                \hline 
                \multirow{2}{*}{Browser} & \multicolumn{2}{c||}{Manifest-Version} &  \multicolumn{3}{c||}{Host Permissions} & \multicolumn{4}{c|}{APIs} \\
                \cline{2-10}
                & V2 & V3 & $<$all\_urls$>$ & \textit{https://*/*} & \textit{http://*/*} & \texttt{Cookies} & \texttt{WebRequest} (WR) & \texttt{WebRequestBlocking} (WRB) & \texttt{DeclarativeNetRequest} (DNR) \\
                \hline
                Firefox & 34,929 & 2,702 & 13,180 & 221 & 57 & 2,154 & 5,219 & 3,626 & 66 \\
                Chrome & 59,020 & 60,132 & 40,363 & 1,452 & 268 & 5,728 & 6,620 & 3,352 & 2,048 \\
                \hline
            \end{tabular}
        }
    \vspace{-1em}
    \label{tab:raw_permsission}
\end{table*}
\begin{table}
\centering
\vspace{-1em}
\caption{Total users affected by each API}
\vspace{-.5em}
        \scalebox{.8}{
            \begin{tabular}{|c||c||c|c||c|}
                \hline
                Browser & \texttt{Cookies} & WR & WRB & DNR \\
                \hline
                Firefox & 13,909,278 & 47,092,558 & 43,306,429 & 70,374  \\
                Chrome & 408,937,955 & 947,610,948 & 489,938,433 & 353,664,323 \\
                \hline
            \end{tabular}
        }
        \label{tab:users}
        \vspace{-1em}
\end{table}
 
\begin{table}
\centering
    \caption{Extensions with "$<$all\_urls$>$" and cookie access}
    \label{tab:hosts_and_perms}
    \vspace{-.5em}
    \scalebox{.8}{
        \begin{tabular}{|c||c||c|c||c|c|}
            \hline
            Browser & \texttt{Cookies} & WR & WRB & DNR & DNR + WR\\
            \hline
            Firefox & 1,409 & 3,440 & 2,415 & 44 & 7\\
            Chrome & 3,274 & 4,278 & 2,032 & 1,391 & 299\\
            \hline
        \end{tabular}
    }
    \vspace{-1em}
\end{table}

 Most modern browsers are implemented on top of a few open-source browsers. Chrome, Microsoft Edge, Opera, and Brave are all built on top of Chromium \cite{chromium}. As such, our extensions worked on all four browsers without modification. We also tested the extensions in the Firefox-based browsers: Firefox and Waterfox. The V2 extensions (\texttt{Cookies} and \texttt{WebRequest}) were directly transferable as Firefox also implements the \texttt{WebExtension} APIs \cite{Needham_2015} to support cross-browser extensions. When using V3, one change was required. Despite defining the host permissions in each extension's manifest, Firefox treats them as optional permissions for V3 extensions \cite{Santala_2022}. Thus, after installation, the extension must prompt the user to re-accept its host permissions. Once granted, V3 extensions behaved as they did in Chromium. We believe that prompting the user does not pose a problem to the adversary, as they could (and likely would) imitate a legitimate service. Another caveat is that Firefox still supports the \texttt{WebRequestBlocking} permission for V3 extensions. Testing in Waterfox obtained the same results.

\subsection{Extension API Permissions' Landscape}
\label{sec:the_scrapening}

The above examples highlight how certain extension APIs can be leveraged to steal and modify cookies (including session cookies). To determine the prevalence of these permissions we crawled Chrome's and Firefox's extension repositories: the Chrome Web Store \cite{chrome_web_store} and \textit{addons.mozilla.org} (AMO) \cite{firefox_amo}. As of March 2024, we identified 122,951 Chrome and 37,783 Firefox extensions. However, of the extensions identified, we could only pull the manifests of 119,152 Chrome and 37,631 Firefox extensions for analysis. Appendix \ref{sec:scraping_appendix} discusses how we scraped each extension in greater detail. We also identified several Chrome Apps and Android extensions. As Chrome Apps are deprecated on all systems except ChromeOS (where the deprecation is ongoing) \cite{chrome_apps} and only 22 extensions were marked as Andriod, we exclude them from our main analysis. Instead, we analyze the discovered Apps and Android variants in Appendix \ref{sec:app_appendix}. All extensions and manifests analyzed are available in our public repository \cite{anonymous642_cream}.

We parsed each extension's manifest for three attributes. First, we obtained the manifest version from the \texttt{manifest\_version} tag. Next, we checked the \texttt{permissions} entry to find if the extension used any of the three APIs discussed in Section \ref{sec:case_study}. Finally, we searched the manifest for the strings  ``$<$all\_urls$>$'', ``*://*/*'', ``https://*/*'', and ``http://*/*''. These strings allow an extension to access every page (first two strings), all HTTPS pages, and all HTTP pages in the browser, respectively \cite{chrome_ext_host_2012}.  While the \texttt{host\_permissions} key lists the sites an extension interacts with, the entry is optional \cite{chrome_host_permissons, mdn_host_permissions}. \texttt{DeclarativeNetRequest} and content scripts can specify hosts within their own manifest entries. Thus an extension without a \texttt{host\_permissions} entry may still have access to cookies. For this reason, we search the whole manifest for the above strings rather than just the \texttt{host\_permissions} entry. Detecting these host permissions in an extension indicates that at least a portion of it has this access. Similarly, not all extensions with ``$<$all\_urls$>$'' behavior use this tag. As such, we treat ``$<$all\_urls$>$'', ``*://*/*'', and the combination of ``https://*/*'' and ``http://*/*'' as equivalent. Table \ref{tab:raw_permsission} summarizes our results. We also calculate the number of users affected by each API as reported by each extension store (depicted in Table \ref{tab:users}).

We found that 59.9\% of extensions are written using manifest V2. However, V2 extensions are much more prevalent in Firefox accounting for 92.8\% of all Firefox extensions. Chrome extensions are more evenly split with 49.5\% still using the V2 format. This disparity is likely due to Chrome's (at the time of testing proposed and now deployed) deprecation of the V2 format \cite{Li_2023} whereas Firefox will continue to support both versions \cite{Sullivan_2024}.

For host permissions, we found that 35.2\% of extensions analyzed define broad host permissions. Unlike manifest versioning, this ratio is consistent across Firefox (35.6\%) and Chrome (35.3\%). Nearly all of these extensions have the ``$<$all\_urls$>$'' behavior. This accounts for 34.2\%, 35\%, and 33.9\% of all, Firefox, and Chrome extensions, respectively. 0.2\% of extensions declared permission for all HTTP sites only and 1\% declared permission for all HTTPS sites alone.

7,882 extensions (5\% of extensions analyzed) used the \texttt{Cookies} API, reaching a combined user base of 422,847,233 users. Note that this combined user base represents an upper bound, as a single user may have multiple extensions and thus be counted multiple times. Similarly, it is unclear whether extension repositories measure users by installations or actual user accounts. More extensions used the \texttt{WebRequest} API. Specifically, 11,839 (7.6\%) were able to monitor network traffic and 6,978 extensions (4.5\%) also declared the \texttt{WebRequestBlocking} permission and could edit HTTP headers. Again while limited extensions declare these permissions, they have nearly one billion users (994,703,506) with 533,244,862 users simultaneously affected by the \texttt{WebRequestBlocking} permission. The \texttt{DeclarativeNetRequest} API is the least used, only appearing in 2,114 extensions (1.3\%). Similarly, it is almost non-existent in Firefox appearing in only 66 extensions (0.17\% of all Firefox extensions). This low adoption is likely due to the transition from V2 to V3 formatting (at least in Chrome), as the \texttt{DeclarativeNetRequest} API is only available to V3 extensions. In that regard, it appeared in 3.4\% of V3 extensions found and has a combined 353,734,697 users.

Each API is limited by the extension's host permissions. Hence, we also examined how many of the above extensions also have ``$<$all\_urls$>$'' host permissions. Table~\ref{tab:hosts_and_perms} summarizes these findings. 54.8\% of extensions with \texttt{Cookies} API permissions have access to all cookies in the browser. Similarly, 65.2\% of \texttt{WebRequest} extensions and 63.7\% of \texttt{WebRequestBlocking} extensions can read and modify any requests in the browser. 67.9\% of manifests with \texttt{DeclarativeNetRequest} permissions can also modify any network requests. Further, 21.3\% of these extensions (14.5\% all declarative extensions) also contain \texttt{WebRequest} permissions, allowing them to view any request (despite contradicting the privacy-preserving philosophy of the declarative API).

Our analysis shows that widespread extensions can access 
users' cookies. Similarly, these extensions have a combined user base in the order of hundreds of millions. Thus if compromised (or malicious), these extensions pose a significant threat to cookies and session security.

\section{\acroc Overview}
\label{sec:nutshell}

Motivated by the results presented in Section~\ref{sec:malicousness}, we advocate for browser controls to allow selective prevention or detection (when prevention is not possible due to application requirements) of cookie theft and illegal cookie modifications made by extensions. To realize these goals we propose \acroc (\underline{C}ookie \underline{R}estrictions for \underline{E}xtension \underline{A}buse \underline{M}itigation) as a modification to the browser's trust model.
This Section focuses on \acroc's concepts and protocol. We discuss browser support for \acroc in Section~\ref{sec:details}.

\begin{figure*}
    \centering
    \scalebox{.9}{
        \scriptsize
        \begin{tabular}{|p{.45\textwidth} p{.17\textwidth} p{.4\textwidth}|}
            \hline
            \textbf{Server} & & \textbf{Browser} \\
            \hline
            \multicolumn{3}{|c|}{\textbf{(A) CREATING A \texttt{Monitored} COOKIE}} \\
            & & 1). $K_{browser} \leftarrow KeyGen()\qquad$\\
            & & 2). Attach $K_{browser}$ to outbound request \\
            \multicolumn{3}{|c|}{\sendmessageleft{top={HTTPS Req. including $K_{browser}$}}} \\
            3). $msg = Enc((Name, Value, K_{browser},...), K_{server})$ & & \\
            \hspace{2em}($K_{server}$ is a secret known only by the server) & & \\
            \multicolumn{3}{|c|}{\sendmessageright{top ={Set-Cookie: $Name=Value;Monitored=msg, ...$}}} \\
            & & 4). Store the received cookie \\
            \hline
            \multicolumn{3}{|c|}{\textbf{(B) SERVER RECEIVING A \texttt{Monitored} COOKIE}} \\
            & & 1). Fetch cookies for the domain \\
            & & 2). If cookie $C_i$ has the \texttt{Monitored} attribute:\\
            & & \hspace{2em} $Report_{C_i} = \{log_{C_i}, Settings_{C_i}, msg_{C_i},timestamp\}$ \\ 
            & & \hspace{2em} H $ = HMAC(Report_{C_i}, K_{browser})$\\
            & & 3). Attach $Report_{C_i}$ and H to the outbound request \\
            \multicolumn{3}{|c|}{\sendmessageleft{top={Request including $C_i$, H, $Report_{C_i}$}}} \\
            4). $Name_{exp}, Value_{exp}, K_{browser}, ... \leftarrow Dec( msg_{C_i}, K_{server})$ & & \\
            5). Check: $Name_{C_i} == Name_{exp} ~\&\&~ Value_{C_i} == Value_{exp}$ & & \\
            6). $Verify(Report_{C_i}, H, K_{browser})$ & & \\
            7). Use $Report_{C_i}$ to validate $C_i$ (per server policy)\vspace{.1em}  & & \\
            \hline
        \end{tabular}
    }
    \caption{Protocols for creation and usage of \texttt{Monitored} cookies}
    \vspace{-1em}
    \label{fig:tracked}
\end{figure*}

\subsection{Threat Model} 
\label{sec:threat_model}

We extend the active network attacker model described in prior work \cite{chen2018we, roth2022security, barth2010protecting}. The model assumes an Adversary (\adv) can intercept, modify, and inject HTTP traffic on the network. It also assumes \adv controls a malicious website. In addition, we assume that \adv controls a malicious extension in the user's browser with arbitrary API permissions and can run on all pages in the browser (e.g. has the ``$<$all\_urls$>$" host permission). As such, \adv can also inject content scripts (malicious JS) into all pages. \adv's goal is to modify or steal a user's cookies through attacks such as cookie hijacking and cookie fixation and gain access to the user's session. 

We assume that the browser core and the Network Service utility process are trusted and vulnerability-free, as \acroc functionality is implemented as part of these modules. Similarly, we trust the underlying host's OS which is relied upon to implement all browser isolation mechanisms (irrespective of \acroc). Attacks that compromise/modify privileged software (e.g., the OS) or the browser software implementation are considered orthogonal and out of scope in this paper. 

\subsection{Rationale}

\acroc introduces two new cookie attributes: \texttt{BrowserOnly} and \texttt{Monitored}. \texttt{BrowserOnly} is a 1-byte flag indicating whether a cookie should be visible to extensions. If set, the cookie is only accessible to the browser. Therefore neither JS nor extensions can access these cookies. \texttt{BrowserOnly} also prevents accesses by JS to ensure an extension cannot access cookies through a content script. The \texttt{BrowserOnly} attribute only protects cookies within the browser. To prevent \texttt{BrowserOnly} cookies from being stolen on the network they should also be marked \texttt{Secure}, which is standard.
 
For cases where cookies must be accessible for websites and extensions to function properly (where \texttt{BrowserOnly} can not be used), we define the \texttt{Monitored} attribute. The \texttt{Monitored} attribute holds a server-defined message that is stored by and only accessible to the browser. The browser also logs all changes made to \texttt{Monitored} cookies. Later, when sending a \texttt{Monitored} cookie to the server,  the browser attaches an authenticated report to the request that contains the server-defined message, the cookie's current settings, and any logged changes. With this, the server can validate the changes made to the cookie and detect if it is stolen.

Since \texttt{Monitored} cookies must be accessible, we cannot prevent them from being stolen. Instead, we make stolen cookies unusable (e.g., for authentication). Figure \ref{fig:tracked} depicts the protocol to create (part {\bf(A)}) and use (part {\bf(B)}) \texttt{Monitored} cookies. The browser generates a symmetric key for each site it visits (step A.1 in Figure \ref{fig:tracked}) and attaches it to all outbound HTTPS requests for that site (step A.2). Note that in this context, a site refers to the protocol and registry controlled domain name of a resource \cite{reis2019site, chromium_site_iso} (e.g., \textit{https://example.com}).
The generated key allows a site to identify the browser in subsequent communication. However, by creating a new key per site the keys cannot be used to track users across sites. When creating a \texttt{Monitored} cookie, the server builds the \texttt{Monitored} attribute's message. This message must include the cookie's name, value, and the received browser key. The cookie's name-value pair ties the attribute to the specific cookie while the browser key ties the whole cookie to a single browser. The server can also include any other data necessary for later cookie validation. The server then uses a secret to symmetrically encrypt the message and uses the resulting ciphertext as the \texttt{Monitored} attribute's value (step A.3). This authenticated encryption ensures the attribute was generated by the server, is only readable to the server, and cannot be altered.

\begin{figure*}[]
    \centering
    \subfigure[Outbound Request]{
        \includegraphics[width=0.47\textwidth]{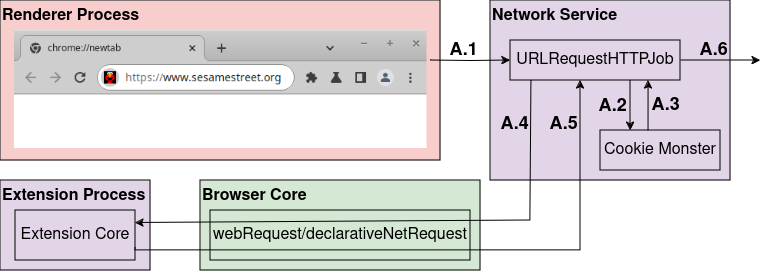}
        \label{subfig:network_request}
    }
    \hfill
    \subfigure[Inbound Response]{
        \includegraphics[width=0.47\textwidth]{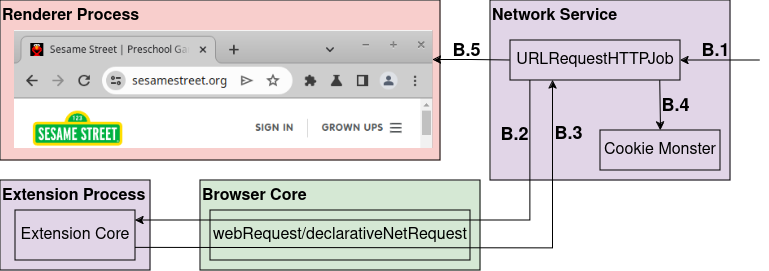}
        \label{subfig:network_response}
    }
    \vspace{-.5em}
    \caption{The flow of requests and responses are through the Network Service}
    \vspace{-1em}
    \label{}
\end{figure*}

When sending the \texttt{Monitored} cookie back to the server, the browser builds its corresponding \texttt{Monitored} report, including an HMAC produced with the browser's key for that site, and attaches it to the request (steps B.1-B.3). Upon receipt, the server decrypts the \texttt{Monitored} message with its secret key (B.4), validates that it is the correct message for the received cookie (B.5), and verifies the HMAC using the key stored in the \texttt{Monitored} attribute (B.6). If the cookie was stolen extension, the HMAC verification would fail as \adv would not know the browser's key (as long as the browser is itself trusted -- recall Section \ref{sec:threat_model}). Similarly, attempts to change the server-defined \texttt{Monitored} message would cause the decryption/validation to fail. As a result, \texttt{Monitored} cookies can only be used by the legitimate browser.

As detection relies on the symmetric key sent to the server in the initial request, the key can only be sent over secure (HTTPS) channels. This ensures the key is always encrypted and inaccessible to \adv over the network. Otherwise, if not encrypted \adv could steal the key, forge authenticated \texttt{Monitored} reports, and use stolen \texttt{Monitored} cookies. However, \texttt{Monitored} cookies themselves can be sent over any channel as the \texttt{Monitored} attribute is always encrypted.

By design, \texttt{Monitored} cookies are editable. As such, the browser logs all changes made to \texttt{Monitored} cookies. This \texttt{changelog} allows the server to validate these changes and take action accordingly (depending on the server's policy). While the \texttt{changelog} stores all changes made to the cookie, it only exists while the cookie exists. Therefore, if \adv (in the form of an extension) deleted the cookie and made a new cookie shadowing the original \cite{squarcina2023cookie} except for \adv intended changes, these changes would not be logged as no \texttt{changelog} exists at this point. Similarly, only edits made through the Cookie Monster are logged thus changes made directly to cookie headers are not. Therefore, when sending the \texttt{Monitored} report to the server, \acroc includes the current value of the cookie's attributes (i.e. its current settings) as well as the \texttt{changelog}. By comparing these current settings to some expected value, the server can detect changes that do not appear in the \texttt{changelog}. The server-authenticated \texttt{Monitored} attribute can store the expected settings if needed. Finally, \acroc also adds a timestamp in the report to prevent replay attacks. As stated above, the entire report is authenticated and sent to the server alongside the \texttt{Monitored} cookie.

Leveraging these controls requires servers to update their behavior, adding the attributes to cookie definitions and, in the case of \texttt{Monitored}, implementing the discussed protocol accordingly. However, these attributes do not alter existing cookie attributes and behavior. Thus only servers that wish to use the \texttt{BrowserOnly} or \texttt{Monitored} attributes must alter their behavior, preserving backward compatibility.

\section{\acroc Architectural Design}\label{sec:details}

As discussed above, the \texttt{BrowserOnly} and \texttt{Monitored} attributes require architectural support from the browser to restrict/monitor extensions' actions. Aside from creating the new cookie attributes, this also requires support from the Cookie Monster and Network Service. As each attribute functions independently, we discuss their required support separately.


\subsection{\texttt{BrowserOnly} Attribute}
\label{sec:browser_only_implementation}

As discussed in Section \ref{sec:nutshell}, \texttt{BrowserOnly} is a 1-byte flag that prevents a cookie from being accessed by JS and extensions. Thus the Cookie Monster must be updated to filter on this attribute. Recall from Section~\ref{sec:cookie_monster} that there are two ways to fetch cookies from the Cookie Monster, \texttt{GetAll} and \texttt{GetList}, and that both \texttt{GetList} and \texttt{Set} filter cookies based on a supplied set of options. As such, we create a new option to exclude \texttt{BrowserOnly} cookies from \texttt{GetList} and \texttt{Set} calls. This option prevents \texttt{BrowserOnly} cookies from being returned or overwritten in most cases. However, as \texttt{GetAll} never filters its results, \texttt{BrowserOnly} cookies are still returned. To address this, we create a new endpoint (\texttt{GetAllExt}) that returns all non-\texttt{BrowserOnly} cookies in the Cookie Monster. These changes are then propagated to the Cookie Manager API and its respective implementations. Finally, the \texttt{Cookies} extension API is updated to use \texttt{GetAllExt} instead of \texttt{GetAll} and the option to exclude \texttt{BrowserOnly} is added to the necessary accesses. This blocks all extension and JS accesses to \texttt{BrowserOnly} cookies through the Cookie Monster.
Note that cookie deletion in response to storage restrictions (``purges'') can be conditioned on cookies' priorities and security attributes \cite{chromium_cookie_code}. Therefore, we also configure the Cookie Monster to not evict \texttt{BrowserOnly} cookies in these scenarios, thus mitigating eviction attacks \cite{squarcina2023cookie, zheng2015cookies}.

While the above changes make \texttt{BrowserOnly} cookies inaccessible to the \texttt{Cookies} API (and JS), they are still exposed to the \texttt{WebRequest} and \texttt{DeclarativeNetRequest} APIs. Protecting \texttt{BrowserOnly} cookies from these APIs requires updating how cookies are handled in network traffic. Figure \ref{subfig:network_request} illustrates how the Network Service builds network requests. When navigating to a website, the Network Service creates a \texttt{URLRequestHTTPJob} to handle the request (step A.1 in Figure \ref{subfig:network_request}). When building the request, the \texttt{HTTPJob} fetches the destination's cookies from the Cookie Monster and adds them to the \texttt{Cookie} request header (A.2 \& A.3). When all the request headers have been created, they are passed to extensions with the above APIs for further processing (A.4). Once all extensions have finished, the headers are returned to the \texttt{HTTPJob} (A.5) and the request is sent to the server (A.6). To hide \texttt{BrowserOnly} cookies from extensions, we withhold them from the \texttt{Cookie} header until after the extensions have executed (A.5). Once the request headers are returned, we add the \texttt{BrowserOnly} cookies to the front of the \texttt{Cookie} header. This ensures the \texttt{BrowserOnly} cookies are parsed first and prevents cookie shadowing attacks \cite{squarcina2023cookie}.

Similarly, Figure \ref{subfig:network_response} displays how the Network Service parses the server's response. When the response is received (step B.1 in Figure \ref{subfig:network_response}), the \texttt{URLRequestHTTPJob} forwards the response headers to extensions (B.2). Once the extensions finish, the headers are returned to the \texttt{HTTPJob} (B.3). At this point, any new cookies are added to the Cookie Monster (B.4) and the response is sent to the renderer for display (B.5). To hide \texttt{BrowserOnly} cookies in responses, we strip all \texttt{Set-Cookie} headers with the \texttt{BrowserOnly} attribute from the response on arrival (B.1). Then when the headers are returned from the extensions (B.3) we reattach the \texttt{BrowserOnly} headers to the response.

For simplicity, we omit some details of how the Network Service handles transactions and interacts with extensions. Requests pass through several classes between the renderer and the network. We focus on the \texttt{URLRequestHTTPJob} as all \texttt{BrowserOnly} (and \texttt{Monitored}) functionality was added to this class. Similarly, when building a network request, its headers are sent to extensions up to three times \cite{chrome_web_req_2024}. Protecting \texttt{BrowserOnly} cookies requires adding the cookies to the request only after all extension passes have finished. The same applies to parsing network replies.

\textbf{Remark:} While \texttt{BrowserOnly} functionality could in principle be incorporated into existing security-relevant attributes (e.g., \texttt{HttpOnly}) merging these functions would change the standard cookie usage model. \acroc instead creates a new attribute that can be used independently, preserving existing usage models and backward compatibility.

\subsection{\texttt{Monitored} Attribute}
\label{sec:tracked_implementation}

The \texttt{Monitored} attribute instructs the browser to maintain a server-defined message and a \texttt{changelog} for the cookie. These values are only accessible to the Network Service. As such, \texttt{Monitored} only requires updating the Cookie Monster's \texttt{Set} behavior to log changes made to \texttt{Monitored} cookies. As discussed in Section \ref{sec:cookie_monster}, \texttt{Set} takes a new cookie, checks if it already exists, and replaces the existing copy (if found). For \texttt{Monitored} cookies, if a preexisting cookie is found, we check whether the new cookie has a \texttt{Monitored} attribute. If the new cookie does not have the \texttt{Monitored} attribute, the original cookie's \texttt{Monitored} message and \texttt{changelog} are copied to the new cookie. Then the two cookies are compared and any differences are added to the newly copied \texttt{changelog}. Otherwise, if the new cookie has a \texttt{Monitored} attribute, it replaces the prior \texttt{Monitored} attribute and the \texttt{changelog} is reset. Similar to \texttt{BrowserOnly}, extensions (and JS) cannot create \texttt{Monitored} cookies. Thus only the server can set a cookie with a \texttt{Monitored} attribute. By keeping the new cookie's attribute the server can update the \texttt{Monitored} attribute as needed. Further, updating the \texttt{Monitored} attribute implies the server has seen the current \texttt{changelog}. Therefore, clearing the \texttt{changelog} removes previously validated entries, minimizing its size. To avoid resource exhaustion attacks \cite{gierlings2023isolated}, a maximum number of \texttt{changelog} entries can be set and a cookie can be flagged as invalid upon reaching this limit.

As discussed in Section \ref{sec:nutshell}, along with the \texttt{Monitored} attribute, the browser generates a symmetric key for each site visited and attaches the key to all HTTPS requests to that site. The browser must maintain a map of keys and their associated sites that is available to \texttt{URLRequestHTTPJobs}. These keys must be created by and maintained within the Network Service. When sending a request, the \texttt{URLRequestHTTPJob} simply fetches the correct key from this map and attaches it to the request using a custom header. To avoid exposure/tampering by extensions, the key is added after extensions have processed the request headers.

When adding \texttt{Monitored} cookies to network requests (or whenever fetched in general), 
the \texttt{changelog} and \texttt{Monitored} message are never included in the cookie. As such, \texttt{Monitored} cookies can be seen by extensions. However, when building the \texttt{Cookie} header, we log which cookies are \texttt{Monitored} cookies. Then, after the extensions have processed the headers, we generate the \texttt{Monitored} report for the \texttt{Monitored} cookies. Specifically, we retrieve each \texttt{Monitored} cookie's \texttt{changelog}, current settings, and \texttt{Monitored} message from the Cookie Monster and add them to the report. Once all the cookies have been added, the report is timestamped and authenticated with the generated key for the site. The report is authenticated as a whole rather than per cookie's sub-report as all cookies for a single site use the same key. Nonetheless, each cookie contains the key within its authenticated/encrypted \texttt{Monitored} message to ensure that the server always receives (and can verify) the key regardless of the cookies present.

Handling \texttt{Monitored} cookies on incoming responses requires no additional changes. Unlike \texttt{BrowserOnly} cookies, \texttt{Monitored} cookies can be seen by extensions. Similarly, even if extensions see the \texttt{Monitored} message, they cannot change it or use it outside the user's browser. At best, an extension can remove the attribute and disable the \texttt{changelog}. However, this would only cause the server to reject the cookie as it would not contain the expected report.

\section{Security Analysis}

We argue that the \texttt{BrowserOnly} and \texttt{Monitored} attributes protect cookies against malicious extensions, including those with highly privileged API permissions (analyzed in Section \ref{sec:case_study}). We argue each attribute's security separately, as \texttt{BrowserOnly} and \texttt{Monitored} are independent.

\subsection{\texttt{BrowserOnly}}

To access a user's session, \adv could try to get or set cookies using the \texttt{Cookies} API. However, \texttt{BrowserOnly} cookies are not editable from and never returned to the \texttt{Cookies} API. Thus, \adv cannot use an extension to access \texttt{BrowserOnly} cookies via the Cookie Monster. 

\adv can also attempt to access cookies in network headers using the \texttt{WebRequest} and \texttt{DeclarativeNetRequest} APIs. Specifically, \adv can target the \texttt{Cookie} and \texttt{Set-Cookie} HTTP headers. When building network requests, \acroc's Network Service withholds \texttt{BrowserOnly} cookies from the request until after extensions have finished executing. Similarly, when parsing replies, the Network Service strips any \texttt{Set-Cookie} headers with the \texttt{BrowserOnly} attribute before passing the response to extensions. Therefore, an extension can never access or edit \texttt{BrowserOnly} cookies in network traffic.

\adv could forge a malicious cookie/header in the network traffic to mimic a \texttt{BrowserOnly} cookie. However when attaching \texttt{BrowserOnly} cookies to network requests the Network Service attaches them to the front of the header. This ensures the correct cookies are always parsed first. When parsing server responses, the Network Service overwrites any malicious headers when reattaching \texttt{BrowserOnly} cookies to the response. \adv could also attempt an eviction attack. The browser limits how many cookies a website can set. If the website goes over this limit, the Cookie Monster automatically deletes older cookies to make space for the new ones. Using the \texttt{Cookies} API \adv could generate hundreds of cookies to trick the Cookie Monster into deleting the \texttt{BrowserOnly} cookie. Then \adv could safely attach a fake \texttt{BrowserOnly} cookie to the request headers directly. However, \acroc configures the Cookie Monster to never purge \texttt{BrowserOnly} cookies for space. Hence, \adv cannot forge cookies or headers to edit \texttt{BrowserOnly} cookies and \adv cannot use a malicious extension to steal or modify \texttt{BrowserOnly} cookies. 

With a content script, \adv can try to access cookies through the webpage. However, the Cookie Monster also excludes \texttt{BrowserOnly} cookies from JS requests.

Finally, an in-network \adv cannot steal or modify \texttt{BrowserOnly} cookies due to TLS, so long as the cookie is also marked \texttt{Secure}.

\subsection{\texttt{Monitored}}

Similar to the cases above, \adv may try to get and set cookies using the \texttt{Cookies} API. While \texttt{Monitored} cookies are accessible, the \texttt{Monitored} message and \texttt{changelog} are never returned to the \texttt{Cookies} API. This information is only available through the report generation endpoint internal to the Network Service. Regardless of whether the \texttt{Monitored} message is available, \adv cannot generate a valid report thus invalidating the cookie. \adv does not know the server's secret key to forge a malicious \texttt{Monitored} message. \adv could try to interleave the \texttt{Monitored} attribute of a different cookie instead, however, the name-value pair in the attribute would not match the stolen cookie. Similarly, \adv cannot change the expected name-value pair due to the authenticated encryption. Even with the correct \texttt{Monitored} message, \adv would not have the browser's key and thus cannot generate a valid authenticated \texttt{Monitored} report. \adv also cannot alter the expected key in the \texttt{Monitored} message as they again do not have the server's secret key.

All changes to a \texttt{Monitored} cookie are appended to the cookie's \texttt{changelog} and sent to the server as part of the authenticated report. Thus, any changes made by \adv are detectable. To avoid detection, \adv must tamper with the \texttt{changelog}. \adv cannot directly remove entries from the \texttt{changelog} as it is append-only. However, the \texttt{changelog} can be cleared if the cookie's \texttt{Monitored} message is updated or the cookie is deleted and remade (thus making a new \texttt{changelog}). The first option is impossible as extensions cannot create \texttt{Monitored} cookies. In the second case, \adv could remake the cookie without the \texttt{Monitored} attribute, however, no report would be generated invalidating the cookie.

\adv could also try to steal and tamper with cookies in network traffic using the \texttt{WebRequest} and \texttt{DeclarativeNetRequest} APIs. \texttt{Monitored} cookies and their messages are accessible through the \texttt{Cookie} and \texttt{Set-Cookie} headers. However, \texttt{Monitored} cookies stolen by \adv are unusable. \adv could also edit \texttt{Monitored} cookies through these headers. Since these edits are done directly to the header, they would not appear in the \texttt{changelog}. As such, editing the \texttt{Set-Cookie} header would alter the cookie's attributes, but these altered settings would appear in the authenticated \texttt{Monitored} report. As a consequence, the server can see that the cookie's settings do not match the expected configuration and detect the changes. Similarly, if \adv tried to revert the changes before the report generation, the reverted changes would be logged and detectable. \adv could also remove the \texttt{Monitored} attribute itself. This would disable the \texttt{changelog}, but no report would be generated, making the cookie invalid. \adv can also edit cookies in the \texttt{Cookie} header, but this header only contains the cookie's name and value. Changing either would invalidate the cookie as it would no longer match the expected name-value pair in the \texttt{Monitored} message. \adv could also attempt to tamper with the \texttt{Monitored} message itself. Attempts to remove, replace, or alter the cookie's \texttt{Monitored} message would invalidate the cookie as it would either generate no report or an invalid report. Finally, \adv could try to tamper with the key and \texttt{Monitored} report also included in network requests. However both the key and report are attached to the request post extensions' accesses (recall  Section~\ref{sec:details}) and thus are not accessible to extensions.

\adv can use a content script to access cookies through JS. Again, while \texttt{Monitored} cookies can be accessed, their \texttt{Monitored} messages and \texttt{changelog}-s cannot. Similarly, any changes made by \adv are appended to the \texttt{Monitored} cookie's \texttt{changelog}. JS cannot access the \texttt{changelog} or create \texttt{Monitored} cookies. Thus any edits by malicious JS to \texttt{Monitored} cookies are detectable.

\adv can also intercept, inject, and modify HTTP traffic. As such, \adv may try to steal and modify \texttt{Monitored} cookies over the network. If the \texttt{Monitored} cookie also has the \texttt{Secure} attribute it is always encrypted on the network and thus inaccessible. If the cookie is not \texttt{Secure}, it may be visible. In these cases, \adv can steal and modify \texttt{Monitored} cookies and their \texttt{Monitored} messages in the same way as a malicious extension. However, just like with a malicious extension, \texttt{Monitored} cookies are not usable outside the user's browser and changes made to network headers invalidate the cookie and are detectable through the \texttt{Monitored} report. Unlike extensions, a network \adv can tamper with the \texttt{Monitored} report as it is included in a custom HTTP header. \adv could strip the report from the request altogether, however, the lack of a report would still invalidate the cookie. \adv could also tamper with the \texttt{changelog} and current settings in the report to hide malicious cookie edits. However, these changes would cause the server-side verification to fail. Similarly, \adv cannot re-authenticate the report after tampering as they do not know the browser's key for the site. Finally, \adv could try to alter the report, authenticate it with a malicious key, and change the expected key in the \texttt{Monitored} message except \adv does not know the server's secret key. As such, \adv cannot re-encrypt the \texttt{Monitored} message, verification will fail, and the cookie will be invalid. \adv could also try to replay a previous valid report to hide their actions, however, the report includes a timestamp to greatly reduce this possibility

Finally, \adv could try to access/change the browser's key attached to network requests. However, these keys are only attached to HTTPS traffic. As such, the key is always encrypted by TLS on the network and inaccessible to \adv.

\section{Implementation \& Evaluation}
\label{sec:evaluation}

We implement \acroc by extending Chromium version 122.0.6168.0 with 535 lines of C++ code across 31 files. For evaluation, we deploy our prototype on an Ubuntu 20.04 machine running at 3.70 GHz with 32GB of RAM. To test \acroc we implemented a simple HTTPS web application using Node.js. The test application allows a user to register an account and log in. When logging in, the server generates a session cookie with pertinent cookie attributes for each test. To allow the server to create cookies with \texttt{BrowserOnly} and \texttt{Monitored} attributes, we added them to a fork of the Node.js cookie library. Our prototype and test network can be found on our open-source repository \cite{anonymous642_cream} while our fork of Node.js's cookie library can be found in a separate repository \cite{anonymous642_cookie}. We reemphasize that usage of these new attributes is optional and \acroc remains fully compatible with websites that choose not to use them.

\subsection{Cookie Sizes \& Browser Storage}
\label{sec:mem_over}

Each attribute requires storing additional cookie data. The \texttt{BrowserOnly} attribute adds one byte of memory to each cookie. The \texttt{Monitored} attribute is larger, requiring 49 additional bytes. This is due to the \texttt{Monitored} message and \texttt{changelog} being a string and vector, respectively. Both are 24-byte long classes (in a 64-bit architecture) plus there is an additional byte for the default empty character in memory. \texttt{Monitored} also relies on symmetric keys. For this, \acroc uses Chromium's built-in AES implementation which uses 256-bit keys.

The \texttt{Monitored} message and \texttt{changelog} entries also affect the size of \texttt{Monitored} cookies. Both fields depend on server-defined attributes, namely the cookie's name, value, domain, and path. Prior studies found the average length of a cookie's name-value pair to be 37 characters \cite{Calvano_2020} and its domain to be 14 characters long \cite{Berry_2012}. We use these averages and a default path of ``/'' for our analysis. In the worst case, all of a cookie's attributes change at once and are included in the resulting \texttt{changelog} entry. \acroc logs changes in an encoded format (i.e. ``SameSite: Strict'' as ``SS:2'') to reduce each entry's size. As such, this would result in a 108-byte entry and a total memory cost of 132 bytes including the string object itself. However, in most cases, we expect fewer attributes to change at once leading to lower log entry sizes. The \texttt{Monitored} message's size also varies on the server's behavior. Every message contains the cookie's name, value, and the browser's key at least. In our tests, the key was 32 bytes long resulting in a 131-byte-long \texttt{Monitored} message.
    
To use either attribute, they must appear in the \texttt{Set-Cookie} header. The \texttt{BrowserOnly} attribute is represented as the 11-byte string ``BrowserOnly'' whereas \texttt{Monitored} includes both ``Monitored='' and the \texttt{Monitored} message. This adds 141 bytes to the header. This is well within Chromium's 1KB cookie attribute limit \cite{awillia_2021}. A key and authenticated report are also attached to network requests for \texttt{Monitored} cookies. The browser's key adds 38 bytes (including the header name) to the request. The attached report contains each \texttt{Monitored} cookie's sub-report, a timestamp, and an HMAC. A cookie's sub-report is its current settings, \texttt{changelog}, and \texttt{Monitored} message. The current settings are a 108-byte string. The \texttt{changelog}'s size depends on its entries. Assuming a single worst-case entry (as before), a cookies sub-report is 252 bytes long (with formatting). The timestamp is 15 bytes long and the HMAC is 32 bytes long. Altogether, the report would be 516 bytes long, though this varies greatly on the number of \texttt{Monitored} cookies and their \texttt{changelog}-s. We discuss potential size reductions in Appendix \ref{sec:future_work}.

\subsection{Operation Run-Times}

\begin{figure}[]
\centering
    \subfigure[Make]{
        \scalebox{.31}{
            \begin{tikzpicture}
                \begin{axis}[
                    xbar,
                    xmin = 0,
                    xlabel = {Time (ms)},
                    symbolic y coords ={B,S,M,BO},
                    ytick = data,
                    axis x line*=left,
                    axis y line*=left,
                    enlarge y limits = .2,
                    label style={font=\LARGE},
                    tick label style={font=\LARGE},
                    legend style={font=\LARGE},
                    bar width = 20pt
                ]
                \addplot [fill=gray] coordinates {
                    (0.412939,B) +- (0.000077904424101611, 0.0)
                    (0.4184195,S) +- (0.0000885965869810835, 0.0)
                    (0.415843,BO) +- (0.0000978768791670586, 0.0)
                    (0.468998,M) +- (0.000234149261697101, 0.0)
                };
                \end{axis}
            \end{tikzpicture}
        }
        \label{fig:enter-label}
    }
    \subfigure[Get]{
        \centering
        \scalebox{.31}{
            \begin{tikzpicture}
                \begin{axis}[
                    xbar,
                    xmin = 0,
                    xlabel = {Time (ms)},
                    symbolic y coords ={B,S,M,BO},
                    ytick = data,
                    axis x line*=left,
                    axis y line*=left,
                    enlarge y limits = .2,
                    label style={font=\LARGE},
                    tick label style={font=\LARGE},
                    legend style={font=\LARGE},
                    legend image post style={scale=1.5},
                    legend image code/.code={
                    \draw [#1] (0cm,-0.1cm) rectangle (0.2cm,0.2cm); },
                ]
                \addplot[fill=darkgray] coordinates {
                    (8.099999905,B) +- (0.00495329439684806, 0.0)
                    (8.199999809,S) +- (0.00502783486707877, 0.0)
                    (0,BO) 
                    (8,M) +- (0.00466815446277482, 0.0)
                };
                \addplot [fill=lightgray] coordinates {
                    (2.699999809,B) +- (0.00278738107442851, 0.0)
                    (2.800000191,S) +- (0.00279465474827471, 0.0)
                    (0,BO)
                    (3,M) +- (0.00125844008445585, 0.0)
                };
                \addplot [fill=gray] coordinates {
                    (0.100128,B) +- (0.0000186263491133426, 0.0)
                    (0.169165,S) +- (0.000053118967305976, 0.0)
                    (0.171578,BO) +- (0.0000284566727150497, 0.0)
                    (0.168195,M) +- (0.0000619276473309349, 0.0)
                };
                \legend{Extension, JS, Net Service} 
                \end{axis}
            \end{tikzpicture}
        }
        \label{fig:enter-label}
    }
    \subfigure[Update]{
        \centering
        \scalebox{.31}{
            \begin{tikzpicture}
                \begin{axis}[
                    xbar,
                    xmin = 0,
                    xlabel = {Time (ms)},
                    symbolic y coords ={B,S,M,BO},
                    ytick = data,
                    axis x line*=left,
                    axis y line*=left,
                    enlarge y limits = .2,
                    label style={font=\LARGE},
                    tick label style={font=\LARGE},
                    legend style={font=\LARGE},
                    legend image post style={scale=1.5},
                    legend image code/.code={
                    \draw [#1] (0cm,-0.1cm) rectangle (0.2cm,0.2cm); }
                ]
                \addplot[fill=darkgray] coordinates {
                    (10.4000001,B) +- (0.005641572335709, 0.0)
                    (10.5,S) +- (0.00454683145053546, 0.0)
                    (0,BO) 
                    (11.4000001,M) +- (0.17460548921399, 0.0)
                };
                \addplot [fill=lightgray] coordinates {
                    (1.400000095,B) +- (0.000805376121671033, 0.0)
                    (1.5,S) +- (0.000819262256506402, 0.0)
                    (0,BO) 
                    (2.399999619,M) +- (0.000987277542288674, 0.0)
                };
                \addplot [fill=gray] coordinates {
                    (0.690705,B) +- (0.0000829890355050697, 0.0)
                    (0.412329,S) +- (0.0000650452501414788, 0.0)
                    (0.6903625,BO) +- (0.00012143062423702139, 0.0)
                    (0.708292,M) +- (0.0000871641478723152, 0.0)
                };
                \legend{Extension, JS, Net Service}  
                \end{axis}
            \end{tikzpicture}
        }
    }
\\{\fbox{\scriptsize BO = BrowserOnly; M=Monitored; S=Standard; B=Baseline.}}
    \caption{Average time to create, get, and update cookies with \acroc vs. unmodified Chromium.}
    \label{fig:cookie_times}
    \vspace{-1em}
\end{figure}

To assess run-time overheads, we measure the time to create, get, and set different cookie types (i.e., ``Standard'' cookies, \texttt{Monitored}, and \texttt{BrowserOnly}) and compare them to an unmodified Chromium instance (``Baseline''). This timing depends greatly on the other processes in the browser. As such, we repeat each operation 100,000 times and report the median completion times (Figure \ref{fig:cookie_times}) to remove the impact of outliers. We measure these times from the Network Service, JS, and extensions as the timing is dependent on where in the browser the operation occurs. As cookies are stored in the Network Service, accesses from the Network Service are generally faster than JS or extensions which require IPC. However, creating cookies and all \texttt{BrowserOnly} tests are only measurable from the Network Service. Similarly, all tests are performed synchronously to ensure accuracy. 

\acroc's additional checks add little overhead to regular cookie operations. Creating cookies without either new attribute was only 5 $\mu$s slower than the baseline. Similarly fetching a cookie from the Network Service, JS, and extensions took 69, 100, and 100 $\mu$s longer than the baseline respectively. Updating a cookie took an additional 100 $\mu$s from both JS and extensions. However, updating a cookie through the Network Service was 278 $\mu$s faster than the baseline. Repeating the experiment several times yielded similar results. As \acroc does not optimize cookie operations, this ``improvement'' likely reflects the browser's impact on execution times rather than an actual reduction. The browser is a large multi-process asynchronous system. As such the execution of other processes in the browser (and underlying host) likely led to the discrepancy seen.

\texttt{BrowserOnly} cookies also present little overhead, requiring 3, 71, and 0 more $\mu$s to be created, retrieved, and updated respectively. \texttt{Monitored} cookies took an additional 56 $\mu$s to create. Fetching a \texttt{Monitored} cookie from the Network Service, JS, and extensions took an additional 68, 300, and $-100$ $\mu$s respectively. Again the improvement seen when fetching through extensions reflects the impact of other processes (and IPC in this case) on execution times. Updating \texttt{Monitored} cookies took 18, 1,000, and 1,000 $\mu$s longer than the baseline respectively. Most of this overhead stems from updating the cookie's \texttt{changelog}. As such, we also test how varying the number of log entries affects a \texttt{Monitored} cookie's update time. For this, we timed how long it took to update the cookie 10 times and found the median to be 7,247 $\mu$s. In the original test, the median time to update a cookie once was 708 $\mu$s. This change reflects a linear increase (some variability due to background browser operations) indicating previous log entries do not impact the update time of subsequent entries.

\subsection{Server-Side}

\begin{figure}[]
\centering
    \scalebox{.34}{
        \begin{tikzpicture}
            \begin{axis}[
                xbar,
                xmin = 5000,
                xlabel = {Replies/Second},
                ytick = {1000,2000,3000,4000,5000},
                xtick = {5000,6000,7000,8000},
                height=300,
                minor x tick num = 1,
                axis x line*=left,
                axis y line*=left,
                enlarge y limits = .15,
                legend style={at={(.95,.8)},anchor=west, font=\Huge},
                legend image post style={scale=1.75},
                label style={font=\LARGE},
                tick label style={font=\LARGE},
                legend style={font=\LARGE},
                legend image code/.code={
                    \draw [#1] (0cm,-0.1cm) rectangle (0.2cm,0.2cm);
                },
                ]
                \addplot [fill=black!20, error bars/.cd, x fixed, x dir=both, x explicit]  coordinates {
                    (7442.0, 1000)  +- (495.1, 0.0)
                    (7565.8, 2000) +- (475.9, 0.0)
                    (7439.5, 3000) +- (447.2, 0.0)
                    (7699.4, 4000) +- (526.3, 0.0)
                    (7581.0, 5000) +- (506.9, 0.0)
                };
                \addplot [fill=gray!80, error bars/.cd, x fixed, x dir=both, x explicit] coordinates {
                    (7488.7, 1000)  +- (465.0, 0.0)
                    (7333.3, 2000) +- (495.8, 0.0)
                    (7710.9, 3000) +- (480.7, 0.0)
                    (7608.1, 4000) +- (491.7, 0.0)
                    (6861.6, 5000) +- (467.9, 0.0)
                };
                \addplot [fill=darkgray!75, error bars/.cd, x fixed, x dir=both, x explicit] coordinates {
                    (6258.5, 1000)  +- (277.0, 0.0)
                    (6212.3, 2000) +- (381.3, 0.0)
                    (6064.4, 3000) +- (412.0, 0.0)
                    (6113.4, 4000) +- (387.3, 0.0)
                    (6086.6, 5000) +- (367.5, 0.0)
                };
                \legend{Baseline, BrowserOnly, Monitored}               
                \end{axis}
        \end{tikzpicture}
        
    }
    \vspace{-.75em}
    \caption{Average server throughput (replies/s) for varying concurrent users}
    \label{fig:throughput}
\vspace{-1em}
\end{figure}

We also measure both attributes' impact on server-side operations. The \texttt{BrowserOnly} attribute defines how cookies are accessed in the browser. Thus \texttt{BrowserOnly} cookies do not add any overhead to the server. \texttt{Monitored} cookies on the hand, change how servers create and parse cookies.
When creating a \texttt{Monitored} cookie, the server must encrypt the corresponding \texttt{Monitored} message. Using AES-256-CBC, this encryption took 6.7 $\mu$s on average. \texttt{Monitored} cookies also add new server-side checks to validate the cookie and report. While the report's MAC is always verified in the same way, cookie validation depends on the server's policy. In our experiments, a cookie was valid if its \texttt{changelog} was empty and current settings matched an expected configuration. For a single user, these checks added 256 $\mu$s to the server-side processing time. To determine the impact of this additional functionality under load, we also measure the cookie validation under various numbers of concurrent users. Using the \texttt{httperf} tool \cite{httperf}, we tested the server with 10, 100, and 1000 concurrent users. In each successive test, the processing time scaled by a factor of 10 (2,600, 28,151, and 298,873 $\mu$s respectively). Thus the \texttt{Monitored} attributes additional server-side overhead scales linearly with the number of concurrent users.

Along with direct processing times, we measure the impact of each attribute on the server's maximum throughput (shown in Figure \ref{fig:throughput}.) Using standard cookies, the server's throughput caps at 7,545 replies/s on average. Varying the number of concurrent connections from 1000 to 5000 had a negligible impact on the server's throughput. \texttt{BrowserOnly} cookies achieved similar throughput in most tests. However, when testing \texttt{BrowserOnly} cookies with 5000 users, the throughput dropped to 6861 replies/s. We believe this decrease reflects variability on the test device itself rather than an actual drop in throughput as a similar decline was not seen between \texttt{Monitored} tests. For \texttt{Monitored} cookies, the server's throughput capped at an average of 6,147 replies/s. This decrease is due to the additional server-side checks performed on \texttt{Monitored} cookies. We note that our throughput results represent a comfortable lower bound, as tests were performed (for the sake of comparison with the baseline) on a single-core experimental Node.js server. Naturally, throughput should increase with server-side resources and load distribution across multiple server-side cores/machines.

\subsection{Latency}

 Finally, we measure the time taken to build requests and parse responses with different cookies. For this, we timed the \texttt{URLRequestHttpJob} behavior as only this portion of the network stack changed. Timing this portion alone also helps reduce noise caused by the rest of the browser. Unlike the previous tests, timing the network stack's behavior required reloading the test page for each sample. As such, we only gathered 10,000 samples to calculate the median in these experiments (depicted in Figure \ref{fig:net_time}).

\begin{figure}[]
\centering
        \scalebox{.35}{
            \begin{tikzpicture}
                \begin{axis}[
                    xbar,
                    xmin = 1,
                    xlabel = {Time (ms)},
                    symbolic y coords ={Baseline,Standard,Monitored,BrowserOnly},
                    ytick = data,
                    axis x line*=left,
                    axis y line*=left,
                    enlarge y limits = .2,
                    legend style={at={(.95,.8)},anchor=west, font=\Huge},
                    legend image post style={scale=1.25},
                    label style={font=\LARGE},
                    tick label style={font=\LARGE},
                    legend style={font=\LARGE},
                    legend image code/.code={
                        \draw [#1] (0cm,-0.1cm) rectangle (0.2cm,0.2cm); },
                    bar width=15pt
                    ]
                    \addplot [] coordinates {
                        (1.712881,Baseline)  +- (0.006004269572, 0.0)
                        (1.768392,Standard) +- (0.006054386499, 0.0)
                        (1.77115,BrowserOnly) +- (0.006104911711, 0.0)
                        (1.932877,Monitored) +- (0.006465334315, 0.0)
                    };
                    \addplot [fill=gray] coordinates {
                        (2.9572685,Baseline) +- (0.007317221505, 0.0)
                        (2.943084,Standard) +- (0.006785782947, 0.0)
                        (3.090255,BrowserOnly) +- (0.007347973352, 0.0)
                        (2.935714,Monitored) +- (0.007174752949, 0.0)
                    };
                    \legend{Request, Response} 
                \end{axis}
            \end{tikzpicture}
        }
    \vspace{-.5em}
    \caption{Latency of parsing network traffic with different types of cookies}
    \label{fig:net_time}
    \vspace{-.5em}
\end{figure}

\begin{figure}
\centering
    \subfigure[\texttt{Chagelog} Entries]{
        \scalebox{.35}{
        \begin{tikzpicture}
            \begin{axis}[
                xbar,
                xmin = 0,
                xlabel = {Time (ms)},
                symbolic y coords ={Empty,1,10,100},
                ytick = data,
                xticklabel style={/pgf/number format/fixed,/pgf/number format/precision=5},
                    scaled x ticks = false,
                axis x line*=left,
                axis y line*=left,
                enlarge y limits = .2,
                label style={font=\LARGE},
                    tick label style={font=\LARGE},
                    legend style={font=\LARGE},
                bar width = 20pt
            ]
            \addplot [fill=gray] coordinates {
                (0.051,Empty)
                (0.052,1)
                (0.051,10)
                (0.200,100)
            };
            \end{axis}
        \end{tikzpicture}
        \label{fig:report_results}
        }
    }
\subfigure[Cookies]{
        \scalebox{.35}{
        \begin{tikzpicture}
            \begin{axis}[
                xbar,
                xmin = 0,
                xlabel = {Time (ms)},
                symbolic y coords ={1,10,50,100},
                ytick = data,
                xticklabel style={/pgf/number format/fixed,/pgf/number format/precision=5},
                    scaled x ticks = false,
                axis x line*=left,
                axis y line*=left,
                enlarge y limits = .2,
                label style={font=\LARGE},
                    tick label style={font=\LARGE},
                    legend style={font=\LARGE},
                bar width = 20pt
            ]
            \addplot [fill=gray] coordinates {
                (0.051,1)
                (0.221,10)
                (1.200,50)
                (2.671,100)
            };
            \end{axis}
        \end{tikzpicture}
        \label{fig:report_results_cookie}
        }
    }
    \vspace{-.5em}
\caption{Latency of generating \texttt{Monitored} reports with different numbers of (A) log entries and (B) cookies}
\vspace{-1em}
\end{figure}

The only change to parsing responses in \acroc is that \texttt{BrowserOnly} cookies must be stripped from responses before extensions execute. As such, responses with a \texttt{BrowserOnly} cookie took 133 $\mu$s (4.5\%) longer than the baseline to parse. Responses with standard or \texttt{Monitored} cookies require no new behavior. Thus parsing either was on par with the baseline, 14 and 21 $\mu$s faster respectively. Again this improvement is due to the variability in other browser processes. Building requests in \acroc requires withholding \texttt{BrowserOnly} cookies until after extensions execute, generating \texttt{Monitored} reports, and adding the browser's key to the headers (HTTPS only). Only using standard cookies, \acroc was 55 $\mu$s (3.2\%) slower than the baseline. Interestingly, attaching the browser's key (the only new behavior in this case) only took 8 $\mu$s on average. Requests with \texttt{BrowserOnly} and \texttt{Monitored} cookies took 58 and 220 $\mu$s (3.4\% and 12.8\%) longer than the baseline respectively.

\texttt{Monitored} cookies' additional overhead
stems from generating the \texttt{Monitored} report. As such we also timed the report generation itself. Specifically, we measure the time to generate a report for a single cookie with various \texttt{changlelog} entries (shown in Figure \ref{fig:report_results}) and multiple cookies (shown in Figure \ref{fig:report_results_cookie}). Generating the report for a cookie with no changes took 50.8 $\mu$s. Reports with 1 and 10 \texttt{changelog} entries added little overhead taking 51.8 and 50.8 $\mu$s respectively. Jumping to 100 changes, the time to generate a report increased to 200.2 $\mu$s. Thus at a certain point, additional \texttt{changelog} entries begin to slow down report generation. However as discussed in Section \ref{sec:tracked_implementation}, cookies with large \texttt{changelog}-s could be marked invalid, preventing them from reaching this threshold. Changing the number of \texttt{Monitored} cookies led to greater variability. Generating a report with 10 \texttt{Monitored} cookies took 4 times longer (221.2 $\mu$s) than with a single cookie. With 50 and 100 \texttt{Monitored} cookies, this time balloons to 1,200 and 2,671 $\mu$s respectively. While large, prior works state that websites have between 10 and 20 cookies on average \cite{munir2023cookiegraph, drakonakis2020cookie, cahn2016empirical}. Therefore we believe these large reports to be unlikely. Further, only \texttt{Monitored} cookies are added to the report. Thus not all of a site's cookies may be included in the report, minimizing this overhead.

\section{Privacy Considerations}
\label{sec:privacy}

CREAM aims to strike a good balance between security and privacy. Below we discuss privacy-related considerations.

{\bf Private/Incognito Browsing.} To limit tracking cookies~\cite{munir2023cookiegraph, mayer2012third, mdn_third_2024}, browsers restrict which contexts cookies can be accessed from. While we discuss the Cookie Monster as a single entity, browsers have different Cookie Monsters for different browsing contexts. For example, Incognito browsing has a separate Cookie Monster from standard browsing. Thus, Incognito cookies are not accessible during standard browsing (and vice-versa). As discussed in Section \ref{sec:cookie_background}, cookies can also be partitioned to restrict their accessibility by embedding context. As \acroc works on individual cookies, it does not inhibit and remains compatible with these privacy features. Further, \acroc-protected cookies remain manageable by users via the normal browser management interfaces.

\textbf{Privacy-Enhancing Extensions.} Some extensions seek to protect user privacy. Among other checks, they often use the \texttt{Cookies} API to remove known tracking cookies from the browser/requests. As the \texttt{BrowserOnly} attribute hides cookies from extensions, servers could leverage it to avoid privacy-enhancing extensions. As a middle ground, it is possible to allow users to selectively disable the \texttt{BrowserOnly} attribute for specific trusted extensions. This opt-out approach would protect the user by default while allowing discretionary disablement. That said, we note that, regardless of the \texttt{BrowserOnly} attribute, user tracking is still possible through techniques such as browser fingerprinting \cite{iqbal2021fingerprinting, pugliese2020long} which are orthogonal to our goal. Further, without approaches to protect cookies, session cookies can be stolen/used, leading to accounts (and all sensitive information therein) being compromised, i.e., a very strong privacy violation. Without browser-based controls such as \acroc, the latter is possible irrespective of privacy-enhancing extensions.

{\bf Tracking and Fingerprinting.} \texttt{Monitored} cookies rely on symmetric keys to tie a cookie to a browser. However, as keys are generated per site, they cannot be used to track the user across domains. It is also possible to generate keys per origin to limit tracking within a single site, e.g., similar to browsers' origin isolation \cite{chromium_origin_iso}. We note that, as session cookies contain the user's authenticated ID, they identify the user. Thus, irrespective of \acroc, multiple session cookies can track the user across sessions. However, it is possible to update/refresh the keys for each session to prevent their use for cross-session tracking. These keys could also allow individual sites to track users across browsing contexts. This can be prevented by generating keys per site and context rather than per site alone. However, this would incur additional runtime and storage costs.

\section{Related Work}
\label{sec:related_work}
Given our focus, this section discusses prior work on cookie security. Additional discussion of prior work on extension security is provided in Appendix \ref{related_extension_work}.

Cookies are vital to the modern web. As such, their adoption, usage, and vulnerabilities have been well-studied \cite{khodayari2022state, drakonakis2020cookie, cahn2016empirical, sivakorn2016cracked, sanchez2021journey}. Their importance has also led to several attempts to mitigate these vulnerabilities. Some prior work introduces additional entities external to the browser to protect session cookies \cite{johns2011reliable, kirda2006noxes, nikiforakis2011sessionshield}. Often using proxies, these controls store session cookies outside the browser to prevent hijacking \cite{nikiforakis2011sessionshield} or associate them with a companion cookie to mitigate fixation attacks \cite{johns2011reliable}. While effective, external storage only works for cookies that are not used within the browser. Additionally, companion cookies require doubling a website's cookies. As browsers limit the number of cookies a single site can make \cite{squarcina2023cookie}, companion cookies effectively half the cookies available to a site.

Session Protection is a similar external control offered by Flask-Login \cite{Countryman_2023} (Flask's session management module). When enabled, the server hashes the user's IP address and user agent (browser) for each session cookie. These hashes are stored on the server and link the cookies to specific hosts. Flask recalculates this hash for all requests with a session cookie and compares it against the saved hash. In the strictest configuration, a hash mismatch invalidates the session. However, by default, a mismatch only marks the session as stale. In this state, the user can still access the site but must re-authenticate for any action that requires a ``fresh'' session. Flask also supports a separate ``Remember Me'' token to save and restore sessions. This token is not protected by Session Protection and thus can be stolen to bypass it altogether. As a consequence, the Flask-Paranoid add-on \cite{Grinberg_2017} extends Session Protection to the Remember Me token as well. Regardless, while Session Protection can detect cookie theft, it cannot accurately determine if a cookie was stolen as benign actions (i.e., IP address changes) also violate the control. Thus stolen cookies could still provide access to most of a session.

Other proposals use extensions to bolster the security of cookies \cite{bugliesi2015cookiext, de2012serene, bugliesi2014provably}. CookieExt \cite{bugliesi2015cookiext} detects session cookies in incoming requests and marks them as \texttt{Secure} and \texttt{HttpOnly}. Serene \cite{de2012serene} maintains a database of session cookie values and ensures the value of outbound cookies are in the database. As discussed in Section \ref{sec:intro}, malicious extensions can bypass existing cookie attributes and reattach cookies to requests. Thus the aforementioned extension-based approaches cannot protect against other extensions. 

Many controls change cookies themselves \cite{bortz2011origin, dacosta2012one, express_signed, google_signed}. Signed cookies contain a cryptographic signature in the cookie's value to support integrity-checking \cite{express_signed} and access-control \cite{google_signed}. Origin cookies \cite{bortz2011origin} add a new ``Origin" attribute that greatly limits a cookie's accessibility to a single origin. One-Time Cookies (OTC) \cite{dacosta2012one} generate a unique token per request rather than per session. OTCs are then tied to a request with a MAC and verified with a shared session key. This session key and any other information needed for verification are stored in a session ticket, encrypted by the server's secret key. These features limit how cookies are accessed, but do not protect against and still allow theft by malicious extensions. While OTCs reduce the usability of a stolen cookie to a single request, they are enabled through HTTP headers. Thus, malicious extensions can disable OTCs outright and force standard cookie authentication.

Finally, similar to \acroc, some approaches implement controls directly in the browser \cite{tang2011fortifying, rfc8471, Monsen_Brigisson_2024}. Origin-Bound Certificates (OBCs) \cite{tang2011fortifying} use on-demand self-signed certificate generation to allow for mutually authenticated TLS connections. Servers can then associate data such as cookies with an OBC tying it to a single device. The Token Binding standard \cite{rfc8471} built upon OBCs, replacing the self-signed certificates with asymmetric keys negotiated during the TLS handshake. The generated Token Binding key and a signature (issued with the corresponding private key) are included in subsequent requests \cite{Anupam_Popov}. The Token Binding key is then used as an ID in cookies to tie them to the channel and device. Chrome also recently proposed Device Bound Session Cookies (DBSC) \cite{Monsen_Brigisson_2024}. DBSC proposes using Trusted Platform Module (TPM) keys to tie cookies to a device. In DBSC, the server creates a TPM key on the user's device. The browser then uses this key to generate a cryptographic proof of identity and sends it to the server. Once verified, the server generates short-lived session cookies to use in future requests.  When these cookies expire, DBSC uses the established key to re-authenticate with the server and get fresh cookies. Similar to the \texttt{Monitored} attribute, these controls prevent cookie theft by linking them to a single device. However, none of them address malicious extensions. In Token Binding, both the cookies and signature are included in request headers. As such, both can be stolen by a malicious extension and used to impersonate the user. Similarly, DBSC only changes how session cookies are generated by the server. Thus once the cookies are generated, an extension can still steal them to gain access to the user's account. DBSC also greatly reduces the lifetime of session cookies, but this only means an adversary needs to steal the cookies more frequently to maintain account access. Also different from \acroc, none of these controls address cookie integrity violations.

\section{Conclusion}

We proposed \acroc: a modification to the web browser's trust model to protect cookies from malicious extensions. \acroc introduces two new cookie attributes: \texttt{BrowserOnly} and \texttt{Monitored}. \texttt{BrowserOnly} prevents cookies from being accessed by extensions and JS. \texttt{Monitored} maintains accessibility but ties the cookie to a single browser, logs all changes made to the cookie, and sends an authenticated report of changes to the server alongside the cookie. This allows the server to detect and refuse stolen and tampered cookies. We provide an open-source prototype of \acroc built atop Chromium \cite{anonymous642_cream} and evaluate it to show that both the \texttt{BrowserOnly} and \texttt{Monitored} attributes incur small overheads.






%



\bibliographystyle{IEEEtran}
\bibliography{TheCount}

\appendices

\section{Scraping Browser Extensions}
\label{sec:scraping_appendix}

For our analysis of current extensions, we used the Chrome Web Store's and AMO's sitemaps \cite{chrome_sitemap, firefox_sitemap} to identify potential extension URLs. In total we found 11,490,897 URLs, however, this contained many repeated URLs and subdomains such as an extension's reviews. Filtering the results, we were left with 172,386 and 533,790 unique Chrome and Firefox extensions respectively. With this, we visited each identified URL and pulled the extension's name, label, and user count from its page. The majority of extensions were labeled ``Themes''. Themes are purely cosmetic and contain no JS or HTML \cite{chrome_themes_2012}. For this reason, we excluded themes from our analysis. Similarly, 16,380 Chrome extensions were unlabeled. Due to this, we believe these extensions to be Chrome Apps, a now deprecated subclass of extension that behaves similarly to native applications \cite{chrome_apps_det_2012}. While Apps are deprecated on most platforms, their deprecation on Enterprise and Education builds of ChromeOS is still ongoing \cite{chrome_apps}. Since Apps are still technically available we do analyze their permissions, however separately from our main discussion. Firefox also labeled 22 extensions as ``Android''. Given this distinction, we also analyze these extensions separately from the main discussion. Filtering out these categories we were left with a dataset of 122,951 Chrome and 37,783 Firefox extensions.

With 177,136 extensions identified (including Chrome Apps and Android variants), the final step was to download each extension's manifest file. For this, we used an online extension viewer \cite{Wu_2023} to pull and decompress each extension's files. Using this tool, we entered each identified extension's URL and downloaded its manifest file. We repeated this process several times. For each successive pass, we rechecked all URLs that resulted in an error with a larger timeout value. This repetition helped remove false errors due to rate limiting on the server and larger extensions requiring more time to load. When complete, we downloaded 119,152 Chrome and 37,631 Firefox extension manifests. The remaining 3,951 extension URLs resulted in various 400 HTTP errors, suggesting they were either private or had been removed from the repository between our two scrapes. We also pulled 12,351 Chrome App and 22 Android manifests.

\begin{table*}[]
    \centering
    \caption{Breakdown of manifest versioning, host permissions, and access to cookies}
    \scalebox{.9}{
        \begin{tabular}{|c||c|c||c|c|c||c|c|c|}
             \hline 
             \multirow{2}{*}{Browser} & \multicolumn{2}{c||}{Manifest-Version} &  \multicolumn{3}{c||}{Host Permissions} & \multicolumn{3}{c|}{APIs} \\
             \cline{2-9}
             & V2 & V3 & $<$all\_urls$>$ & \textit{https://*/*} & \textit{http://*/*} & \texttt{Cookies} & \texttt{WebRequest} & \texttt{WebRequestBlocking} \\
             \hline
             Android & 20 & 2 & 18 & 0 & 0 & 2 & 14 & 13 \\
             Chrome Apps & 12094 & 62 & 455 & 11 & 38 & 15 & 27 & 13 \\
             \hline
        \end{tabular}
    }
    \vspace{-1em}
    \label{tab:bonus_raw_permsission}
\end{table*}

\section{Chrome Apps \& Android Analysis}
\label{sec:app_appendix}

As discussed above, we discovered 16,380 Chrome Apps and 22 Android (Firefox) extensions. From this, we successfully pulled 12,156 App and 22 Android manifests. Similar to our main discussion we analyzed each manifest for its version, host permissions, and access to cookies. A summary of these findings can be found in Table \ref{tab:bonus_raw_permsission}.

Nearly all manifests analyzed (99.5\%) were written using manifest V2. This is likely due to a lack of incentive to change formats as Firefox currently has no plans to deprecate the format and Chrome Apps will be fully deprecated by the end of 2025 \cite{chrome_apps_det_2012}. Looking at host permissions, only 4.3\% of manifests declared broad host permissions. However, most Android extensions (81.8\%) had 
``$<$all\_urls$>$" permissions. Only 4.1\% of Chrome Apps had any broad host permissioning. These limited permissions may be due to Chrome Apps performing more like native applications thus running more locally rather than needing to access many websites.

We saw similar low numbers concerning access to cookies. Only 17 manifests (0.1\%) had the \texttt{Cookies} permission. Similarly, only 41 (0.3\%) used the \texttt{WebRequest} API with 26 (0.2\%) also having access to the \texttt{WebRequestBlocking} permission. While low, 63.6\% of Android extensions had the \texttt{WebRequest} permission, and all but one of these also declared the \texttt{WebRequesBlocking} permission. Despite this, manifests with the \texttt{Cookies}, \texttt{WebRequest}, and \texttt{WebRequestBlocking} API affected a combined 2,245,591, 14,699,240, and 13,622,263 users respectively. No manifest declared the \texttt{DeclarativeNetRequest} permission.

We further analyzed which manifests with access cookies also had ``$<$all\_urls$>$" host permissions. Table \ref{tab:bonus_hosts_and_perms} summarizes these findings. Both Android extensions with the \texttt{Cookies} permission could access all cookies in the browser. This was less common in Chrome Apps. Only 20\% of Apps with \texttt{Cookies} permissions were able to access all cookies. 92.9\% of Android extensions and 48.1\% of Chrome Apps with \texttt{WebRequest} permissions could see all traffic in the browser. Similarly, all of these Android extensions and 60\% of these Chrome Apps also declare the \texttt{WebRequestBlocking} permission.

\begin{table}[]
    \centering
    \caption{Manifests with ''$<$all\_urls$>$'' and cookie access}
    \scalebox{.8}{
    \begin{tabular}{|c|c|c|c|}
        \hline
        Browser & \texttt{Cookies} & \texttt{WebRequest} & \texttt{WebRequestBlocking} \\ 
        \hline
        Android & 2 & 13 & 13 \\
        Chrome Apps & 3 & 13 & 8 \\
        \hline
    \end{tabular}
    }
    \vspace{-1em}
    \label{tab:bonus_hosts_and_perms}
\end{table}

\section{Additional Discussion \& Limitations}
\label{sec:future_work}

\textbf{Reduced Memory Overhead:} As mentioned in Section \ref{sec:mem_over}, both the \texttt{changelog} and \texttt{Monitored} report can grow rapidly for cookies with frequent changes. Thus one direction for future work is to minimize the size of these values. There are several ways this reduction may be achieved. In our design, both the \texttt{Monitored} log entries and messages are C++ string objects. These objects add 24 bytes of overhead to each cookie and each \texttt{changelog} entry. Thus switching to another representation such as a basic character pointer could help reduce some of this storage overhead. Similarly, we already encode the \texttt{changelog} entries and cookie's current settings to minimize their size. However, as they are both strings it may be possible to use other string compression algorithms to reduce their size further. Finally, it may be possible to reduce which attributes are logged. Some attributes such as \texttt{HttpOnly} prevent cookies from being stolen, but a stolen \texttt{Monitored} cookie cannot be used outside the user's browser. As such it may not be necessary to log changes made to these attributes. 

Several possible server-side changes could also reduce the memory overhead of \texttt{Monitored} cookies. Each \texttt{Monitored} cookie's message contains an encrypted copy of the browser's key for the domain. However all \texttt{Monitored} cookies for the same domain share the same key. Thus it may be possible to store the encrypted key in a subset of cookies that cover all possible requests. This would reduce the \texttt{Monitored} message and report size of many \texttt{Monitored} cookies as they would no longer need to store a duplicate encrypted key. Further, if the server stored the browser's key itself (similar to proposals like DBSC \cite{Monsen_Brigisson_2024} or frameworks like Flask \cite{Countryman_2023}) there would be no need for the \texttt{Monitored} message at all. However, this would trade browser overhead for server overhead. Finally, it may be possible to create a new server endpoint (also similar to DBSC) to receive \texttt{Monitored} reports. This would move the reports from a header and into the request body which has fewer size limitations. However, all of these changes would require increased server-side support.

\textbf{Further Extension Analysis:} During our analysis of extensions, we specifically looked for which extensions could access cookies. However, during the process, we noted some other interesting patterns. During the scrape, we identified several nearly identical extensions. These extensions had identical names, highly similar descriptions and assets, and the majority of their code was the same. However, it is unclear why so many duplicates exist and how they differ. Similarly, while Chrome and Firefox both support the \texttt{WebExtension} APIs they do not always share the same implementation. Despite this many extensions work across browsers. Due to these observations, another interesting direction for future work is further analysis of these extensions to determine how they differ from each other, across browsers, and what effect this may have on the security of the browser.

\section{Extension Security: Extended Related Work}
\label{related_extension_work}
Given their popularity and access to powerful APIs, several studies have analyzed extensions in terms of the APIs they commonly use, how they impact other controls, and how users understand their impacts \cite{picazo2022chrome, bui2023detection, agarwal2022helping, kariryaa2021understanding}. Additionally, prior work also investigates the privacy threats of extensions, such as tracking user behavior and interests \cite{karami2020carnus}. As such many works investigate extension fingerprinting techniques \cite{solomos2022dangers,solomos2022escaping} and mitigations \cite{sjosten2019latex,karami2022unleash}.

As mentioned in Section \ref{sec:intro}, extensions can be exploited to yield extension permissions to \adv. As such, many analysis techniques and browser modifications have been proposed to help detect/mitigate extension vulnerabilities \cite{yu2023coco,fass2021doublex,some2019empoweb,kim2023extending,xie2024arcanum}. Several studies also acknowledge the threat posed by malicious extensions focusing on detection \cite{kapravelos2014hulk, xing2015understanding, jagpal2015trends, dekoven2017malicious, aggarwal2018spy, chen2018mystique, pantelaios2020you} whereas fewer prior attempts aim to prevent malicious extension actions. WebEnclave \cite{wang2021webenclave} is an extension that creates an isolated enclave on a website to hide secure elements such as login prompts from other extensions. Similarly, WRIT \cite{vasiliadis2023writ} uses service workers to attach tokens to JS events (i.e. mouse clicks) to verify if they were generated by the user rather than an extension. Picazo-Sanchez et. al. \cite{picazo2020after} create a browser extension to monitor changes made by each extension, remove them before the next extension runs, and re-add all the changes after all extensions have executed. The latter prevents a malicious extension from abusing execution orders to learn about/tamper with other extensions. Unlike \acroc, these works focus on website interactions rather than broader browser APIs like \texttt{Cookies} and \texttt{WebRequest}.


\end{document}